\documentclass[apjl]{emulateapj}

\def\ltsima{$\; \buildrel < \over \sim \;$}
\def\simlt{\lower.5ex\hbox{\ltsima}}
\def\gtsima{$\; \buildrel > \over \sim \;$}
\def\simgt{\lower.5ex\hbox{\gtsima}}

\newcommand{\myemail}{matsuoka@arcetri.astro.it}
\newcommand{\nustar}{{\it NuSTAR}}
\newcommand{\spitzer}{{\it Spitzer}}
\newcommand{\chandra}{{\it Chandra}}
\newcommand{\xmm}{{\it XMM}-{\it Newton}}

\shorttitle{Nature of Hard X-ray Detected Hy/U/LIRGs in the COSMOS Field}
\shortauthors{Matsuoka and Ueda}

\begin{document}

\title{Nature of Hard X-ray (3--24 keV) Detected Luminous Infrared Galaxies\\in the COSMOS Field}

\author{Kenta Matsuoka\altaffilmark{1,2,3,4} and Yoshihiro Ueda\altaffilmark{1}}

\altaffiltext{1}{Department of Astronomy, Kyoto University, Kitashirakawa-Oiwake-cho, Sakyo-ku, Kyoto 606-8502, Japan}
\altaffiltext{2}{Dipartimento di Fisica e Astronomia, Universitˆ di Firenze, Via G. Sansone 1, I-50019 Sesto Fiorentino, Italy}
\altaffiltext{3}{INAF -- Osservatorio Astrofisico di Arcetri, Largo Enrico Fermi 5, I-50125 Firenze, Italy}
\altaffiltext{4}{\myemail}

\begin{abstract}
We investigate the nature of far-infrared (70 \micron) and hard X-ray (3--24 keV) selected galaxies in the COSMOS field detected with both \spitzer\ and Nuclear Spectroscopic Telescope Array (\nustar).
By matching the \spitzer-COSMOS catalog against the \nustar-COSMOS catalog, we obtain a sample consisting of a hyperluminous infrared galaxy with $\log (L_{\rm IR}/L_\odot) \geq 13$, 12 ultraluminous infrared galaxies with $12 \leq \log (L_{\rm IR}/L_\odot) \leq 13$, and 10 luminous infrared galaxies with $11 \leq \log (L_{\rm IR}/L_\odot) \leq 12$, i.e., 23 Hy/U/LIRGs in total.
Using their X-ray hardness ratios, we find that 12 sources are obscured active galactic nuclei (AGNs) with absorption column densities of $N_{\rm H} > 10^{22}$ cm$^{-2}$, including several Compton-thick ($N_{\rm H} \sim 10^{24}$ cm$^{-2}$) AGN candidates.
On the basis of the infrared (60 \micron) and intrinsic X-ray luminosities, we examine the relation between star-formation (SF) and AGN luminosities of the 23 Hy/U/LIRGs.
We find that the correlation is similar to that of optically-selected AGNs reported by \citet{2009MNRAS.399.1907N}, whereas local, far-infrared selected U/LIRGs show higher SF-to-AGN luminosity ratios than the average of our sample.
This result suggests that our Hy/U/LIRGs detected both with \spitzer\ and \nustar\ are likely situated in a transition epoch between AGN-rising and cold-gas diminishing phases in SF-AGN evolutional sequences.
The nature of a Compton-thick AGN candidate newly detected above 8 keV with \nustar\ \citep[ID 245 in][]{2015ApJ...808..185C} is briefly discussed.
\end{abstract}

\keywords{galaxies: active --- galaxies: evolution --- infrared: galaxies}

\section{Introduction}\label{sec:i}

The coevolution of galaxies and supermassive black holes (SMBHs) is one of the most crucial issues in modern extragalactic researches.
This idea originated from observational results of a tight correlation between bulge velocity dispersions ($\sigma$) and black hole masses ($M_{\rm BH}$) in the local universe \citep[e.g.,][]{1998AJ....115.2285M,2003ApJ...589L..21M,2009ApJ...698..198G,2013ApJ...772...49W}.
However, we have no definite scenario how galaxies have coevolved with SMBHs through the cosmic history.
To unveil this mechanism, the galaxy merger would be an important clue.
\citet{1988ApJ...325...74S} proposed that a major-merger evolutionary scenario in which galaxy mergers induce a rapid starburst and an obscured BH growth, followed by an unobscured epoch after the gas is consumed or expelled from the galaxy by a feedback \citep[see Fig.~6 in][]{2012NewAR..56...93A}.
Furthermore, from the theoretical point of view, we have some frameworks that are able to produce the local $M_{\rm BH}$-$\sigma$ relation \citep[e.g.,][]{2006ApJ...641...90R,2011MNRAS.412.2154B}.

In addition to cumulative results, i.e., the $M_{\rm BH}$-$\sigma$ relation, the connection between star formation (SF) and active galactic nucleus (AGN) is also a key phenomenon in revealing the on-going interaction between galaxies and SMBHs.
\citet{2011MNRAS.412.2154B} showed simultaneous bursts of SF and BH growth based on $N$-body simulations of galaxy mergers \citep[see also][]{2010MNRAS.407.1529H}.
Such theoretical results expect a positive correlation between SF and AGN luminosities.
In fact, \citet{2009MNRAS.399.1907N} reported a positive linear trend between SF and AGN luminosities by using local type-2 AGNs and high-$z$ quasars \citep[see also][]{2007ApJ...666..806N,2014ApJ...784..137K,2015ApJ...807...28M}.
However, these studies were mainly based on moderately matured AGNs in the ``unobscured'' epoch.
Furthermore, recent studies have shown that the relation between SF and AGN luminosities of distant X-ray detected AGNs is flat (i.e., there is no significant correlation) if the sample is divided in redshift bins \citep[e.g.,][]{2012A&A...545A..45R,2015MNRAS.453..591S}.
In order to understand the SF-AGN connection correctly, it is crucial to focus on heavily-obscured AGNs, which would show violent star formation with buried BH growth.

In the major-merger scenario, dusty populations, like ultraluminous infrared galaxies (ULIRGs) and submillimeter galaxies (SMGs), are considered to be in an explosive evolutionary phase just after a merger.
Unfortunately, we cannot investigate their AGN activities in optical or ultraviolet (UV) wavelengths since such populations usually conceal their nuclear activities with dense gas.
Although re-emitted lights from heated dust in infrared (IR) wavelength may be used to estimate their AGN activities, separation from SF activities is often non-trivial.
An alternative, very efficient way to unveil the AGN activities is X-ray observations.
Especially, high energy X-ray emission above $\sim 10$ keV allow us to detect even heavily-obscured, even Compton-thick AGNs with absorption column densities of $N_{\rm H} \sim 10^{24}$ cm$^{-2}$, thanks to its strong penetrating power against photoelectric absorption.
Therefore, hard X-ray observations are one of the most suitable tools to investigate the hidden SF-AGN connection in dusty populations.

\begin{deluxetable*}{cccccccc}
\tablecaption{Properties of 23 Hy/U/LIRGs \label{tab:t}}
\tablewidth{0pt}
\tablecolumns{8}
\tablehead{
\colhead{ID$_{\it Sp}$\tablenotemark{$a$}} & \colhead{ID$_{\it Nu}$\tablenotemark{$b$}} &
\colhead{$z$\tablenotemark{$c$}} &
\colhead{$\log (L_{\rm IR}/L_\odot)$\tablenotemark{$d$}} &
\colhead{$\log \lambda L_{\rm 60{\mu}m}$\tablenotemark{$e$}} &
\colhead{$\log N_{\rm H}$\tablenotemark{$f$}} &
\colhead{$\log L_{\rm X}$\tablenotemark{$g$}} &
\colhead{$\log L_{\rm AGN}$\tablenotemark{$h$}}\\
\colhead{} & \colhead{} & \colhead{} & \colhead{} & \colhead{[erg s$^{-1}$]} &
\colhead{[cm$^{-2}$]} & \colhead{[erg s$^{-1}$]} & \colhead{[erg s$^{-1}$]}
}
\startdata
1190 & 103 & 0.520 & 11.83$\pm$0.04 & 45.01$\pm$0.10 & 23.4$_{-3.4}^{+0.5}$ & 44.21$\pm$0.38 & 45.80$\pm$0.54\\
1224 & 107 & 0.694 & 12.09$\pm$0.27 & 45.25$\pm$0.10 & 23.8$_{-1.0}^{+0.3}$ & 44.75$\pm$0.40 & 46.53$\pm$0.51\\
1222 & 111 & 0.500 & 11.65$\pm$0.22 & 45.13$\pm$0.07 & 20.0$_{-0.0}^{+0.0}$ & 44.17$\pm$0.11 & 45.74$\pm$0.43\\
1563 & 123 & 0.787 & 12.02$\pm$0.29 & 45.13$\pm$0.11 & 23.8$_{-1.4}^{+0.3}$ & 44.74$\pm$0.42 & 46.52$\pm$0.51\\
1928 & 145 & 0.445 & 11.29$\pm$0.26 & 44.43$\pm$0.10 & 22.2$_{-0.0}^{+0.1}$ & 43.97$\pm$0.07 & 45.48$\pm$0.27\\
2613 & 188 & 0.560 & 11.94$\pm$0.06 & 45.02$\pm$0.12 & 22.5$_{-0.0}^{+0.1}$ & 44.12$\pm$0.11 & 45.68$\pm$0.43\\
2659 & 192 & 0.360 & 11.61$\pm$0.26 & 44.54$\pm$0.27 & 20.9$_{-0.9}^{+0.3}$ & 43.66$\pm$0.10 & 45.07$\pm$0.38\\
2872 & 194 & 1.156 & 12.53$\pm$0.31 & 45.27$\pm$0.22 & 22.1$_{-0.1}^{+0.0}$ & 45.26$\pm$0.05 & 47.24$\pm$0.20\\
3064 & 206 & 1.024 & 12.36$\pm$0.27 & 45.35$\pm$0.17 & 21.9$_{-0.1}^{+0.0}$ & 45.29$\pm$0.04 & 47.28$\pm$0.16\\
3180 & 216 & 0.760 & 12.14$\pm$0.28 & 45.26$\pm$0.07 & 23.8$_{-0.8}^{+0.2}$ & 44.86$\pm$0.32 & 46.68$\pm$0.43\\
3603 & 232 & 0.931 & 11.98$\pm$0.23 & 45.10$\pm$0.18 & 23.4$_{-0.1}^{+0.1}$ & 44.77$\pm$0.10 & 46.56$\pm$0.35\\
5191 & 245 & 1.250 & 12.07$\pm$0.31 & 46.30$\pm$0.21 & 24.0$_{-4.0}^{+0.5}$ & 45.28$\pm$0.51 & 47.27$\pm$0.51\\
5074 & 251 & 1.066 & 12.75$\pm$0.29 & 45.54$\pm$0.11 & 20.0$_{-0.0}^{+0.0}$ & 44.97$\pm$0.08 & 46.83$\pm$0.31\\
5041 & 253 & 0.212 & 11.07$\pm$0.10 & 44.10$\pm$0.10 & 22.7$_{-0.0}^{+0.1}$ & 43.36$\pm$0.09 & 44.69$\pm$0.34\\
4644 & 256 & 0.260 & 11.38$\pm$0.03 & 44.44$\pm$0.08 & 20.6$_{-0.6}^{+0.6}$ & 43.38$\pm$0.08 & 44.71$\pm$0.30\\
4383 & 287 & 0.658 & 12.05$\pm$0.25 & 45.13$\pm$0.09 & 20.0$_{-0.0}^{+0.0}$ & 44.40$\pm$0.07 & 46.05$\pm$0.27\\
3756 & 296 & 1.108 & 12.36$\pm$0.23 & 45.72$\pm$0.16 & 21.5$_{-0.5}^{+0.3}$ & 44.95$\pm$0.08 & 46.81$\pm$0.31\\
4017 & 307 & 0.345 & 11.29$\pm$0.26 & 44.38$\pm$0.07 & 20.9$_{-0.3}^{+0.1}$ & 43.74$\pm$0.12 & 45.18$\pm$0.46\\
3967 & 311 & 0.688 & 12.37$\pm$0.03 & 45.60$\pm$0.04 & 22.9$_{-0.0}^{+0.1}$ & 44.38$\pm$0.13 & 46.03$\pm$0.51\\
3827 & 320 & 0.690 & 12.07$\pm$0.26 & 45.17$\pm$0.12 & 20.0$_{-0.0}^{+0.0}$ & 44.36$\pm$0.13 & 46.00$\pm$0.50\\
3654 & 322 & 0.356 & 11.45$\pm$0.04 & 44.56$\pm$0.12 & 21.9$_{-0.1}^{+0.0}$ & 43.84$\pm$0.07 & 45.31$\pm$0.27\\
3424 & 337 & 1.454 & 12.87$\pm$0.33 & 45.75$\pm$0.29 & 21.4$_{-1.4}^{+0.3}$ & 45.01$\pm$0.11 & 46.89$\pm$0.43\\
3638 & 339 & 1.845 & 13.03$\pm$0.09 & 46.52$\pm$0.32 & 23.3$_{-0.1}^{+0.0}$ & 45.57$\pm$0.08 & 47.68$\pm$0.32
\enddata
\tablecomments{
All the errors are 1$\sigma$.
}
\tablenotetext{$a$}{Identification number in the \spitzer-COSMOS catalog \citep{2010ApJ...709..572K}.}
\tablenotetext{$b$}{\nustar\ source number from \citet{2015ApJ...808..185C}.}
\tablenotetext{$c$}{Redshift provided by \citet{2010ApJ...709..572K}.}
\tablenotetext{$d$}{IR luminosity integrated from 8 \micron\ to 1000 \micron\ in \citet{2010ApJ...709..572K}.}
\tablenotetext{$e$}{FIR luminosity at 60 \micron\ (rest-frame) estimated by using \spitzer/MIPS data at 24, 70, and 160 \micron\ (observed frame) bands (see Section~\ref{sec:d2}).}
\tablenotetext{$f$}{Absorption hydrogen column density estimated by using the HR value (see Section~\ref{sec:r} for more details). We adopt $\log N_{\rm H} = 20.0$ as a dummy lower limit.}
\tablenotetext{$g$}{Intrinsic (de-absorbed) luminosity in the rest-frame 2--10 keV band.}
\tablenotetext{$h$}{Bolometric AGN luminosity converted from the 2--10 keV luminosity by adopting a conversion factors given in \citet{2009ApJ...700.1878R}.}
\end{deluxetable*}

In this paper, we focus on IR-luminous galaxies at $0 < z < 3$ that are detected in hard X-rays above 3 keV in the COSMOS (Cosmic Evolution Survey) field to reveal the hidden SF-AGN connection.
Using a sample obtained by matching the \spitzer\ 70\micron\ catalog \citep{2010ApJ...709..572K} against the Nuclear Spectroscopic Telescope Array (\nustar) hard X-ray (3--24 keV) catalog \citep{2015ApJ...808..185C}, we investigate their obscuration properties and relation between SF and AGN luminosities.
Our work is unique in using hard X-rays above 8 keV for the first time on this subject at distant universe, and is complementary to \citet{2016MNRAS.456.2735L}, who studied the X-ray properties below 8 keV of the \spitzer\ sources based on the \xmm\ \citep{2010ApJ...716..348B} and \chandra\ \citep{2009ApJS..184..158E} catalogs.
We adopt a concordance cosmology with $(\Omega_M, \Omega_\Lambda) = (0.3,0.7)$ and $H_0 = 70$ km s$^{-1}$ Mpc$^{-1}$.
All the attached errors correspond to 1$\sigma$.

\section{Sample}\label{sec:s}

\begin{figure}
\epsscale{0.995}
\plotone{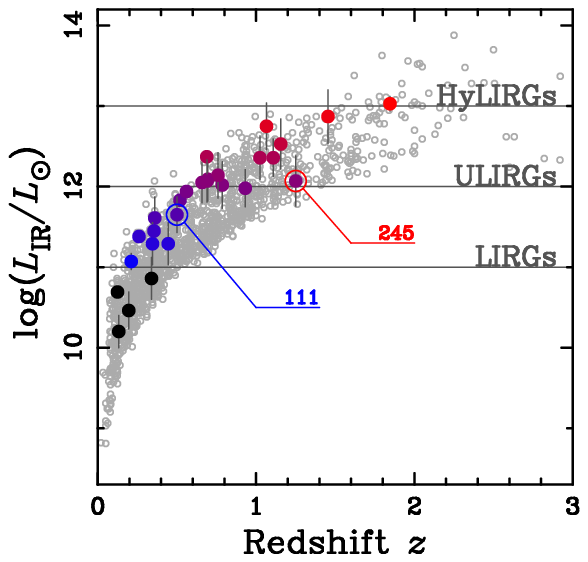}
\caption{Redshift versus IR (8--1000 $\mu$m) luminosity plot of the COSMOS sample.
Filled circles denote \nustar-detected IR galaxies, color-coded according to the $\log(L_{\rm IR}/L_\odot)$ value (bluer and redder for lower and higher IR luminosity objects, respectively).
Objects with $\log(L_{\rm IR}/L_\odot) < 11$ are shown in black.
The remaining undetected objects in \citet{2010ApJ...709..572K} are shown as grey-small circles.
Three horizontal lines represent the dividing luminosities for LIRGs, ULIRGs, and HyLIRGs, corresponding to $\log(L_{\rm IR}/L_\odot) = 11$, 12, and 13, respectively.
\nustar-detected objects without \chandra\ and/or \xmm\ detections \citep[i.e., ID~245 and 111 in][]{2015ApJ...808..185C} are marked with large-open circles in red and blue, respectively.}
\label{fig:s}
\end{figure}

In this study, we utilize a large galaxy sample constructed by \citet{2010ApJ...709..572K} mainly based on the \spitzer-COSMOS survey \citep[S-COSMOS;][]{2007ApJS..172...86S}, a \spitzer\ legacy survey designed to cover the entire COSMOS field with the Multiband Imaging Photometer on \spitzer\ \citep[MIPS;][]{2004ApJS..154...25R}.
The MIPS data were obtained with 5$\sigma$ sensitivities of 0.08, 8.5, and 65 mJy at 24, 70, and 160 \micron, respectively.
By using the deep MIPS imaging data and multiwavelength dataset of the COSMOS, \citet{2010ApJ...709..572K} obtained 1503 unconfused 70\micron-selected sources.
Thanks to the wide area and deep photometries in the COSMOS, they cover a wide IR luminosity ($L_{\rm IR}$, integrated from 8 to 1000 \micron) range from $10^9$ to $10^{14}$ in units of solar luminosity at $0 < z < 3$ (grey-open circles in Figure~\ref{fig:s}).
Note that 602 objects of them ($\sim$ 40\%) have spectroscopic redshifts.

In order to obtain hard X-ray properties of these IR galaxies, we match them against the \nustar-COSMOS catalog \citep{2015ApJ...808..185C}.
The \nustar\ catalog contains 91 sources detected in the 3--24, 3--8, and/or 8--24 keV bands with sensitivities of 5.9, 2.9, and $6.4 \times 10^{-14}$ erg cm$^{-2}$ s$^{-1}$, respectively.
The survey covers $\approx 1.7$ deg$^2$ of the COSMOS field, overlapping both with the \chandra\ and \xmm\ data.
Four out of the 91 sources are associated with neither \chandra\ nor \xmm\ point-like counterparts.
To best estimate the \nustar-source positions, we basically refer to the coordinates of the \chandra\ counterparts, or to those of the \xmm\ ones in the case of no \chandra\ detection.
Matching radii of 1\arcsec and 3\arcsec are adopted for \chandra\ and \xmm, respectively \citep[e.g.,][]{2012ApJS..201...30C,2007ApJS..172..353B}.
Regarding the four sources without \chandra\ or \xmm\ counterparts, we conservatively adopt matching radius of 25\arcsec, which is smaller than that (30\arcsec) used by \citet{2015ApJ...808..185C}, in order to reduce the probability of chance coincidence.
As the result, we obtain 27 \nustar-detected IR galaxies denoted with filled circles in Figure~\ref{fig:s}, including two objects detected only with \nustar.
In this paper, we focus on 23 sources at $\log (L_{\rm IR}/L_\odot) \geq 11$, since we aim to investigate IR-luminous galaxies.
Among them, 18 objects (78\%) have spectroscopic redshifts.
The averaged number density of 70 \micron\ sources at $\log (L_{\rm
IR}/L_\odot) \geq 11$ is 514 deg$^{-2}$, and thus the expected number
of \spitzer\ sources found by chance around 
a \nustar\ source in 1\arcsec, 3\arcsec, and 25\arcsec\ radii is 0.00012, 0.0011, and 0.078, respectively.
The contamination fraction is small enough to discuss the overall statistical properties of our sample even for the sources detected only with \nustar.

\begin{figure}
\epsscale{1.000}
\plotone{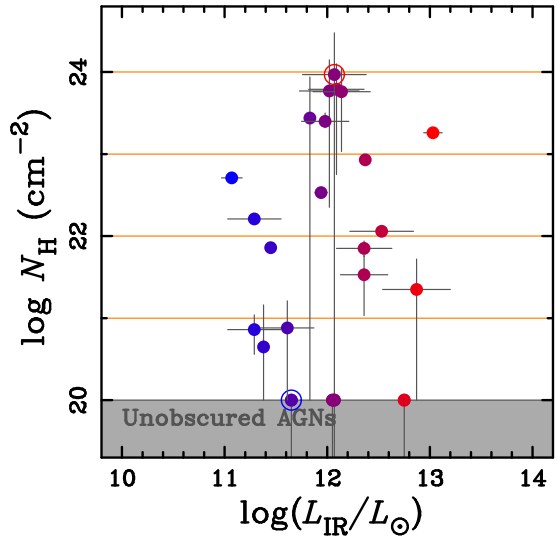}
\caption{Absorption hydrogen column-density versus 
IR (8--1000 $\mu$m) luminosity 
plot for the Hy/U/LIRGs, shown as filled circles color-coded according to the $\log (L_{\rm IR}/L_\odot)$ value (same as Figure~\ref{fig:s}).
Four orange horizontal lines denote column densities of $N_{\rm H} = 10^{21}$, $10^{22}$, $10^{23}$, and $10^{24}$ cm$^{-2}$.
Unobscured objects are plotted on an arbitrary line at $\log N_{\rm H} = 20$.
Sources ID~245 and 111 are marked with red and blue open circles.}
\label{fig:n}
\end{figure}

\section{Results}\label{sec:r}

Our sample contains a hyperluminous IR galaxy, 12 ULIRGs, and 10 luminous IR galaxies, i.e., 23 Hy/U/LIRGs, divided by criteria of $\log (L_{\rm IR}/L_\odot) \geq 13$, $12 \leq \log (L_{\rm IR}/L_\odot) < 13$, and $11 \leq \log (L_{\rm IR}/L_\odot) < 12$, respectively.
Source properties (e.g., redshift and IR luminosity) are listed in Table~\ref{tab:t}.
The \nustar\ detection rate in the 70\micron-selected sample at $\log (L_{\rm IR}/L_\odot) \geq 11$ is 2.3\%; those in each subsample are 3.9\% for HyLIRGs, 4.7\% for ULIRGs, and 1.7\% for LIRGs.
This trend seems to be inconsistent with previous studies that more IR-luminous galaxies show higher AGN fractions \citep[e.g.,][]{1995ApJS...98..171V,1999ApJ...522..113V,2010ApJ...709..884Y,2014ApJ...794..139I}, although these numbers strongly depend on \nustar\ detection limits.

As described in Section~\ref{sec:s}, two of our \nustar-detected IR galaxies have no lower energy X-ray counterparts detected with \chandra\ and/or \xmm\ \citep[i.e., sources ID~111 and ID~245 in][]{2015ApJ...808..185C}, which are marked with blue and red large open circles in Figure~\ref{fig:s}, respectively.
These may be either variable objects and/or heavily obscured AGNs only detectable in hard X-rays above 8 keV.
In our matching, the former source is assigned as a LIRG at a photometric redshift of 0.5 with an offset of $\approx 20''$ from the \nustar\ position, although \citet{2015ApJ...808..185C} mentioned that this source is coincident with an extended \chandra\ and \xmm\ source at $z = 0.220$ located about 24\arcsec\ from the center.
The latter source (ID~245) was reported as two bright spectroscopically identified galaxies at $z = 1.277$ and 0.358 with positional offsets of 11\arcsec\ and 15\arcsec\ in \citet{2015ApJ...808..185C}, respectively.
By using the IR selection criterion, the object at $z = 1.277$ is identified as an AGN in \citet{2012ApJ...748..142D}. \citet{2015ApJ...808..185C} confirmed that 
the $z=1.277$ object is the counterpart of the \nustar\ source ID~245 by checking the relation between 12 \micron\ and intrinsic X-ray luminosities \citep[e.g.,][]{2009ApJ...693..447F,2009A&A...502..457G}.
Our matching result indicates that this source is a ULIRG at a photometric redshift of 1.25 with an offset of $\approx$ 20\arcsec\ from the \nustar\ position, which is probably the counterpart of the $z=1.277$ object in \citet{2012ApJ...748..142D}.

\begin{figure*}
\epsscale{0.938}
\plottwo{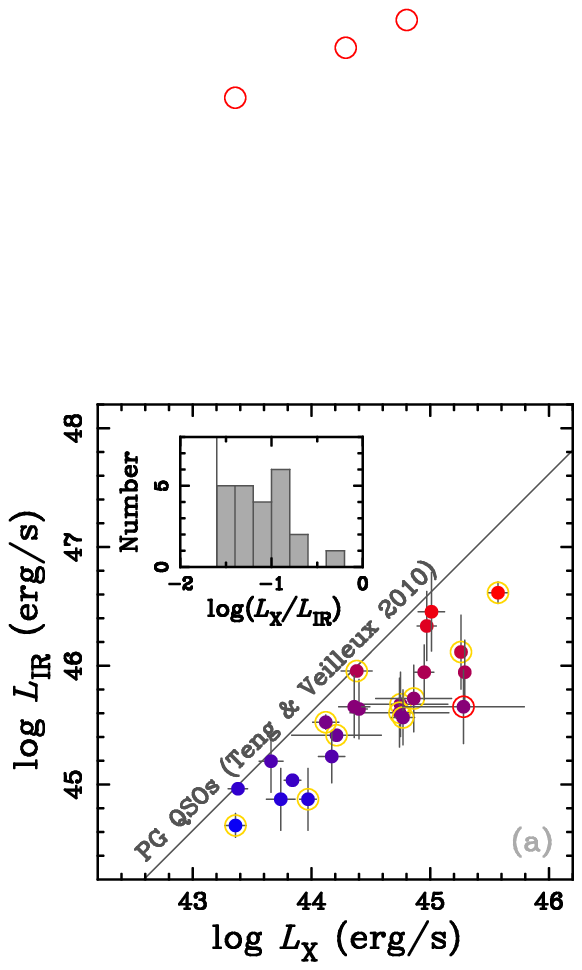}{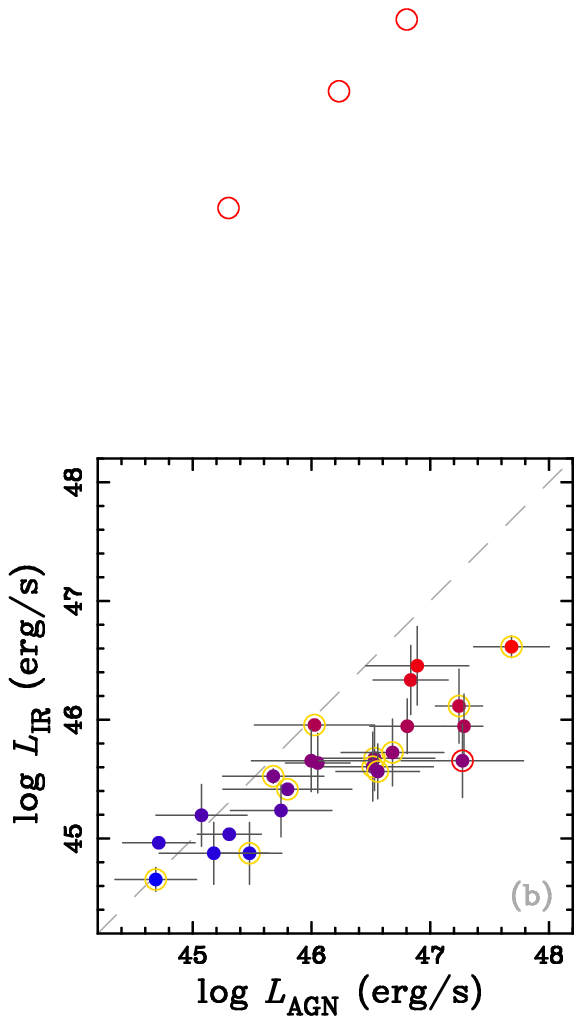}
\caption{(a) Relation between IR (8--1000 $\mu$m) and 2--10 keV luminosities of our Hy/U/LIRGs, shown as filled circles color-coded according to the $\log (L_{\rm IR}/L_\odot)$ value (same as Figure~\ref{fig:s}).
Source ID~245 is marked with a red-open circle.
Obscured objects with $\log N_{\rm H} > 22$ are marked with yellow-open circles.
The grey line denotes the average relation for PG QSOs \citep{2010ApJ...725.1848T}.
The small panel shows the histogram of $\log (L_{\rm X}/L_{\rm IR})$ of our sample.
(b) Relation between IR (8--1000 $\mu$m)
and AGN bolometric luminosities of our Hy/U/LIRGs. The symbols are the same as (a). 
The grey dashed line corresponds to the $L_{\rm IR} = L_{\rm AGN}$ relation.}
\label{fig:t}
\end{figure*}

To examine X-ray obscuration of the \nustar-detected Hy/U/LIRGs, firstly we utilize the hardness ratio defined by \citet{2015ApJ...808..185C}, HR $=(H-S)/(H+S)$, where $H$ and $S$ are vignetting-corrected count rates in the 8--24 keV and 3--8 keV bands, respectively.
In our sample, three out of 23 sources are not detected in the 8--24 keV band and hence have only upper limits of HR.
Using the HR value, we estimate the corresponding absorption hydrogen column density at the source redshift ($N_{\rm H}$).
Following \citet{2003ApJ...598..886U}, we assume a cutoff power law with a photon index of $\Gamma = 1.9$ and an e-folding energy of 300 keV, plus its reflection component from cold matter with a solid angle of $2\pi$ (calculated with the \textbf{pexrav} model in XSPEC), as the intrinsic spectrum, both of which are subject to same absorption.
An unabsorbed scattered component with a fraction of 1\%, a median value of hard X-ray selected local AGNs \citep{2016ApJS..225...14K}, is also added, whose photon index is linked to the intrinsic one ($\Gamma=1.9$).
To make the model applicable to Compton-thick AGNs, attenuation of the transmitted component by Compton scattering (with the \textbf{cabs} model in XSPEC) is considered.
The \nustar\ HR cannot well constrain the column densities of lightly obscured (e.g., $N_{\rm H} \simlt 10^{22}$ cm$^{-2}$) objects, which show low energy cutoff only below $\simlt$ 3 keV.
Thus, for objects whose column density is consistent with zero from the \nustar\ HR within the error, we redetermine it utilizing the \chandra\ HR between the 2--7 keV and 0.5--2 keV count rates based on \citet{2009ApJS..184..158E}, or the \xmm\ HR between the 2--10 keV and 0.5--2 keV count rates based on \citet{2009A&A...497..635C} if the \chandra\ data are unavailable.
Figure~\ref{fig:n} plots the estimated column densities of our sample.
From the count rate and $N_{\rm H}$, we also calculate the intrinsic (de-absorbed) luminosity in the rest-frame 2--10 keV band ($L_{\rm X}$).
The estimated $N_{\rm H}$ and $L_{\rm X}$ values are tabulated in Table~\ref{tab:t}.

As shown in Figure~\ref{fig:n}, on the basis of the best-fit $N_{\rm H}$, we find that 12 (7) out of the 23 Hy/U/LIRGs are consistent with being obscured (heavily obscured) objects with $N_{\rm H} \geq 10^{22}$ (10$^{23}$) cm$^{-2}$, but please acknowledge the uncertainty of this result given to the large errors in $N_{\rm H}$.
Notably, several objects including source ID~245 are Compton-thick AGN candidates whose absorptions reach $\log N_{\rm H} \sim 24$ within the uncertainties.
The remaining four objects indicate no absorption and are plotted on an arbitrary line ($\log N_{\rm H} = 20$) in Figure~\ref{fig:n}.
Because source ID~111 apparently shows no absorption, we infer that this object would be a variable AGN that was fainter in the \chandra\ or \xmm\ observations than in the \nustar\ ones.
Another possibility is that ID~111 is a heavily obscured AGN but contaminated by the soft, extended source at $z=0.22$ detected with \chandra\ and \xmm\ \citep{2015ApJ...808..185C}.

\section{Discussion}\label{sec:d}

Using the \nustar\ and \spitzer\ catalogs in the COSMOS field, we have constructed a new Hy/U/LIRG sample at $z \sim$ 0--3
detected in both the hard X-ray band in the 3--24 keV and the far-infrared (FIR) band.
They are expected to be a key population in understanding the co-evolution of galaxies and SMBHs when the Universe was the most active.
On the basis of a relation between their IR and X-ray luminosities, we examine a hidden connection between SF and AGN activities in these galaxies.

\subsection{Relations between IR and AGN Luminosities}\label{sec:d1}

Figures~\ref{fig:t}(a) and (b) plot the relations between AGN and IR (8--1000 $\mu$m) luminosities of our Hy/U/LIRGs.
In Figure~\ref{fig:t}(a) we use the absorption-corrected 2--10 keV AGN luminosities ($L_{\rm X}$) derived in Section~\ref{sec:r}, while in Figure~\ref{fig:t}(b) we have converted them into bolometric AGN luminosities ($L_{\rm AGN}$) by adopting luminosity-dependent conversion factors ($L_{\rm AGN}/L_{\rm X}$) given by \citet{2009ApJ...700.1878R}.
As shown in Figure~\ref{fig:t}(a), the X-ray to IR luminosity ratio of our sample is $\log (L_{\rm X}/L_{\rm IR}) \sim -1$ with a scatter of $\sim 0.5$ dex.
These ratios are much larger than those of the majority of local ULIRGs \citep[$-4.5 < \log (L_{\rm X}/L_{\rm IR}) <-1.5$;][]{2010ApJ...725.1848T}, and are even larger than the average value of PG QSOs \citep[$\log (L_{\rm X}/L_{\rm IR}) \sim -1.5$;][]{2010ApJ...725.1848T} plotted as the grey line in Figure~\ref{fig:t}(a).
This indicates that our Hy/U/LIRG sample are very ``X-ray luminous'' populations, which are different from typical local ULIRGs.
In Figure~\ref{fig:t}(b), we see that the relation between $L_{\rm IR}$ and $L_{\rm AGN}$ flattens at high $L_{\rm AGN}$, although the trend is not evident in Figure~\ref{fig:t}(a).
This is mainly because the bolometric correction factor in \citet{2009ApJ...700.1878R} rapidly increases with $L_{\rm X}$.

The infrared luminosity in the 8--1000 $\mu$m band should contain contributions both from relatively cool and hot dust from the SF and AGN activities, respectively, even though these $L_{\rm IR}$ values were derived by \citet{2010ApJ...709..572K} on the basis of spectral energy distribution (SED) fit with star-forming galaxy templates. Particularly in luminous AGNs, AGN components can significantly contribute to the IR SED even at $\lambda > 8$ \micron\ \citep[e.g.,][]{2003A&A...403..119B,2008A&A...484..631V}.
Using a 3D Monte-Carlo radiative transfer code with a two-phase clumpy dusty-torus model, \citet{2016MNRAS.458.2288S} calculated the ratio between the predicted IR 1--100 $\mu$m luminosity from the torus and the intrinsic AGN luminosity, $L_{\rm torus}/L_{\rm AGN}$, as a function of torus covering factor. According to the AGN population synthesis model by \citet{2014ApJ...786..104U}, the torus covering factor at $z=1$ is 0.65 and 0.46 for $\log L_{\rm AGN}=46$ and $\log L_{\rm AGN}=47$, respectively (see \citealt{2016MNRAS.458.2288S}).
In moderately optically-thick tori ($\tau = 5$ at 9.7 $\mu$m), \citet{2016MNRAS.458.2288S} predict that the $L_{\rm torus}/L_{\rm AGN}$ ratio is 0.6 (0.2) and 0.3 (0.1) at $\log L_{\rm AGN} = 46$ and $\log L_{\rm AGN} = 47$, respectively, for type-1 (type-2) AGNs.
Considering that about half of $L_{\rm torus}$ is emitted at $\lambda > 8$ \micron\ \citep{2016MNRAS.458.2288S}, the observed $L_{\rm IR}/L_{\rm AGN}$ values of the Hy/U/LIRGs with $\log L_{\rm AGN} \sim 47$ are reasonably consistent with these predictions (i.e., $L_{\rm IR}/L_{\rm AGN} \sim$ 0.05--0.15).
This means that the AGN contribution to the 8--1000 $\mu$m luminosity is significant in these X-ray luminous objects.
Accurate separation of the SF and AGN contributions and study of the broadband AGN spectrum requires a detailed analysis of the IR SED, which is beyond the scope of this paper.
In the next subsection, we use the FIR luminosity at 60 \micron\ to roughly estimate the SF contribution.

\subsection{Relation between SF and AGN Luminosities}\label{sec:d2}

\begin{figure}
\epsscale{0.988}
\plotone{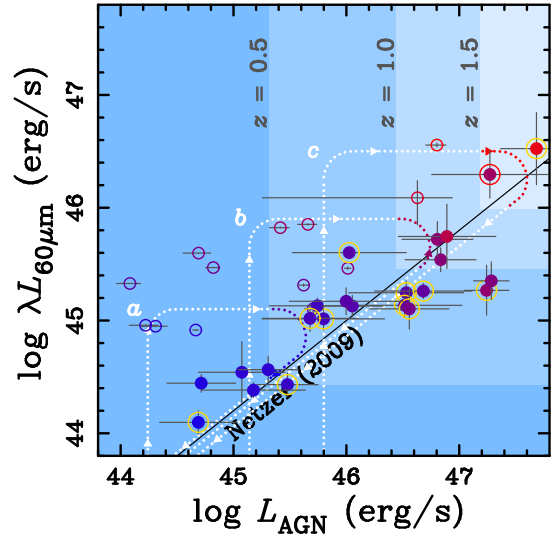}
\caption{Relation between 60 \micron\ and AGN bolometric luminosities of the 23 Hy/U/LIRGs, shown as filled circles color-coded according to the $\log (L_{\rm IR}/L_\odot)$ value (same as Figure~\ref{fig:s}).
Yellow-open circles indicate objects with $\log N_{\rm H} > 22$.
The black line denotes a relation for unobscured AGNs obtained by \citet{2009MNRAS.399.1907N}.
Three rectangle fields marked with $z = 0.5$, 1.0, and 1.5 denote the detectable area for objects with $N_{\rm H} = 10^{23}$ cm$^{-2}$ at each redshift above the \nustar\ and \spitzer\ detection limits.
Three dotted curves denote SF-AGN evolution sequences (see text).
Local U/LIRGs \citep{2014ApJ...794..139I} are shown with open circles color-coded according to IR luminosity.}
\label{fig:l}
\end{figure}

In this subsection, we investigate a relation between FIR and AGN bolometric luminosities of our Hy/U/LIRGs to understand their SF-AGN link.
As a reliable indicator of SF luminosities, we derive FIR luminosities at 60 \micron\ (source-frame) by inter/extrapolating their IR spectral energy distributions obtained with \spitzer/MIPS at 24, 70, and 160 \micron\ bands (observed frame) compiled by \citet{2010ApJ...709..572K}.
Figure~\ref{fig:l} plots the relation between 60 \micron\ and AGN luminosities of our sample.
As noticed, there is a positive correlation between $L_{\rm 60\mu m}$ and $L_{\rm AGN}$.
To compare our result with previous studies of optical-selected AGNs, we also plot the $L_{\rm SF}$-$L_{\rm AGN}$ relation of \citet{2009MNRAS.399.1907N} in the black line, which was obtained from local Seyfert 2 galaxies and quasars at $2 < z < 3$.
We find that the correlation of our Hy/U/LIRGs is similar to that of these optically-selected AGNs.

\citet{2009MNRAS.399.1907N} suggested a SF-AGN evolution sequence that is divided into three phases, i.e., a long star-forming phase, an AGN-rising phase, and a cold-gas diminishing phase.
In the first phase, the SF activity increases with a low AGN luminosity.
This appears as a shift from bottom to top on the $L_{\rm SF}$-$L_{\rm AGN}$ plane.
Next, this SF activity causes more fuelling of cold gas onto the central SMBH, and the AGN activity is enhanced by keeping the high SF luminosity.
This phase corresponds to a shift from left to right in the plot.
As cold gas is consumed with SF and accretion onto the SMBH, both SF and AGN luminosities gradually decrease along the unobscured-AGN relation.
We show these schematic sequences as three dotted lines in Figure~\ref{fig:l}.

According to this scenario, a majority of IR-luminous galaxies would be located above the unobscured-AGN relation of \citet{2009MNRAS.399.1907N}.
In fact, \citet{2014ApJ...794..139I} reported, on the basis of IR spectroscopy, that local U/LIRGs exhibit higher SF luminosities than optically selected Seyfert galaxies with the same AGN luminosities, as shown in open circles in Figure~\ref{fig:l}.
Their result supports that U/LIRGs are situated at the SF-dominated epoch in a SF-AGN evolutionary scenario.
By contrast, however, most of our Hy/U/LIRGs follow the same relation of \citet{2009MNRAS.399.1907N} and no such SF-dominant objects are detected on the top-left side of the black line in Figure~\ref{fig:l}.

We suggest that this apparent inconsistency can be explained by selection effects of our sample.
In Figure~\ref{fig:l}, we show the sensitivity limits in terms of 60 \micron\ and AGN luminosities at $z = 0.5$, 1.0, and 1.5, with blue-colored rectangle areas.
Here we adopt the MIPS 5$\sigma$ sensitivities and the \nustar\ detection limit, $5.9 \times 10^{-14}$ erg cm$^{-2}$ s$^{-1}$ in the 3--24 keV band, by assuming an absorption of $\log N_{\rm H} = 23$.
It is seen that the ``cross points'' of the two sensitivities in the FIR and X-ray bands are {\it coincidently} located close to the \citet{2009MNRAS.399.1907N} line.
For example, the evolutional track $a$ in Figure~\ref{fig:l} denotes an evolution of a SF-dominant LIRG at $z = 0.5$.
Because of the \nustar\ detection limit, $\log L_{\rm AGN} = 45.3$ at $z = 0.5$, we cannot detect LIRGs with fainter X-ray luminosities: undetectable parts are plotted with white dotted lines.
Similarly, evolutional tracks for ULIRGs at $z = 1.0$ and HyLIRGs at $z = 1.5$ are shown as tracks $b$ and $c$, respectively.
On the other hand, while more IR-luminous galaxies with the same AGN luminosities are detectable in terms of the FIR sensitivity, their number density should be smaller and hence it is more difficult to find them within a limited survey volume.
Due to these reasons, we are not able to find SF-dominant IR-luminous galaxies as reported by \citet{2014ApJ...794..139I} in our sample.
Note that such observational limits can produce an artificial correlation between the AGN and SF luminosities.
In order to reveal whether the $L_{\rm AGN}$-$L_{\rm SF}$ relation is real or not, we need X-ray and IR data with sufficient depths and areas.

As we mentioned in Section~\ref{sec:r}, source ID~245 has no soft X-ray counterpart with \chandra\ and \xmm, and shows a high column density of $\log N_{\rm H} \sim 24$.
Thus, this is a Compton-thick AGN candidate that was detectable only by using hard X-rays above 8 keV.
Although the IR luminosity of this object is relatively low (see Figure~\ref{fig:t}), the 60 \micron\ luminosity is not that low, i.e., $\log (\lambda L_{\rm 60 \mu m}/L_{\rm IR}) = 0.64\pm0.37$.
This ratio seems to be higher than the mean ratio ($-0.42\pm0.21$) of the rest of objects, all of which show $\log (\lambda L_{\rm 60 \mu m}/L_{\rm IR}) < 0$.
Note that source ID~245 is detected with MIPS only in the observed frame 24 and 70 \micron\ bands, and only has a loose upper limit on the 160 \micron\ flux, which is important to constrain the peak luminosity of the FIR dust components around (rest frame) 60 \micron.
Thus, there might be large uncertainties in the estimates of $L_{\rm 60 \mu m}$ (based on extrapolation from the 24 and 70 \micron\ fluxes) and $L_{\rm IR}$ (based on SED fit).
Nevertheless, we point out that the slope in the rest frame 11--31 \micron\ range is very steep, $\log (L_{\rm 31 \mu m}/L_{\rm 11 \mu m}) \approx 1.2$, implying that cool dust is abundant in this galaxy.
Therefore, we claim that source ID~245 would be an obscured AGN with violent star formation, in which a huge amount of galactic gas and dust is responsible for the obscuration.

Our result indicates that at least some populations of Hy/U/LIRGs are distributed on the same relation of unobscured populations \citep{2009MNRAS.399.1907N} on the $L_{\rm FIR}$ vs. $L_{\rm AGN}$ plane.
We suggest that these Hy/U/LIRGs are possibly in a transition phase where obscured AGNs have been just unveiled to become unobscured ones, corresponding to a stage between the AGN-rising and cold-gas diminishing phases in the SF-AGN evolution scenario.
This evolutional stage would be important to understand the negative feedback from AGN to SF activities.

\section{Conclusion}

To understand the hidden SF-AGN connection of obscured AGNs, we matched
\spitzer-selected galaxies against \nustar\ sources in the COSMOS field.
As the result, we obtain 23 Hy/U/LIRGs (i.e., a HyLIRG,
12 ULIRGs, and 10 LIRGs) with hard X-ray (3--24 keV) detections.  By
using their X-ray hardness ratios, we estimate their absorption column
densities and intrinsic AGN luminosities.  Then, we investigated their
relation between SF and AGN luminosities.  The main results are
summarized as follows.

\begin{enumerate}
\item We found that our Hy/U/LIRGs are intrinsically quite X-ray
      luminous with respect to the IR (8--1000 \micron) luminosity,
      $\log (L_{\rm X}/L_{\rm IR}) \sim -1$, as compared with local
      ULIRGs and PG QSOs. The contribution from AGN-heated hot dust to
      the IR luminosity is significant in the luminous AGNs.

\item We found an apparent positive trend between SF
and AGN luminosities of our Hy/U/LIRGs, which is similar to that of
optically-selected AGNs \citep{2009MNRAS.399.1907N}.  This implies that
our X-ray detected Hy/U/LIRGs are likely in a transition phase between
obscured and unobscured AGNs.

\end{enumerate}

To investigate heavily-obscured Hy/U/LIRGs in a SF-rising epoch at
distant universe, which should be crucial to understand the whole
picture of SF-AGN relation, deeper hard X-ray survey data covering above
8 keV would be required.

\acknowledgments

We would like to thank Taiki Kawamuro, Atsushi Tanimoto, Kohei Ichikawa, and Tohru Nagao for helpful comments and suggestions, and Takamitsu Miyaji for his help on the energy responses of \chandra\ and \xmm.
We also thank the anonymous referee for useful comments and suggestions.
This work was financially supported by the Japan Society for the Promotion of Science (JSPS) KAKENHI Grant No. 14J01811 (K.M.) and 26400228 (Y.U.).

\appendix

\section{Error and Modification}

\begin{figure*}[h]
\epsscale{0.938}
\plottwo{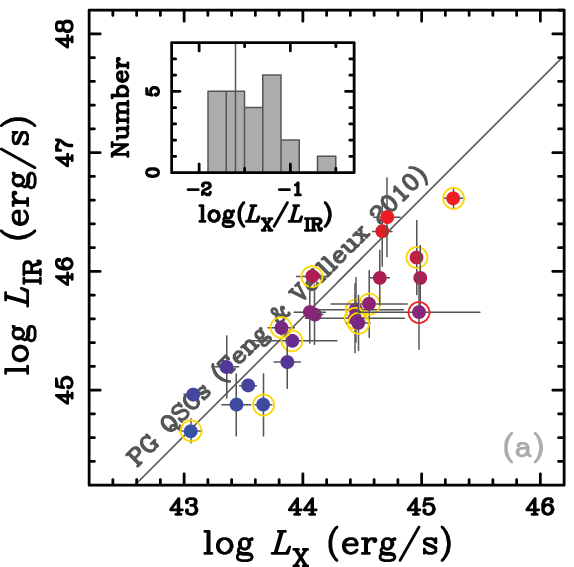}{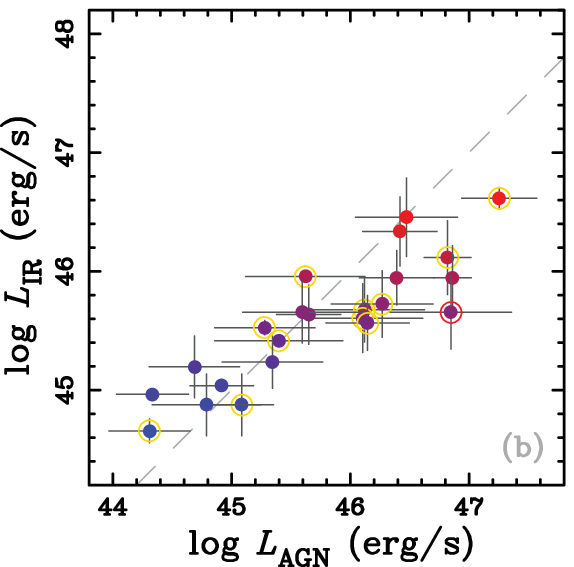}
\caption{(a) Relation between IR (8--1000 $\mu$m) and 2--10 keV luminosities of our Hy/U/LIRGs, shown as filled circles color-coded according to the $\log (L_{\rm IR}/L_\odot)$ value.
Source ID~245 is marked with a red-open circle.
Obscured objects with $\log N_{\rm H} > 22$ are marked with yellow-open circles.
The grey line denotes the average relation for PG QSOs \citep{2010ApJ...725.1848T}.
The small panel shows the histogram of $\log (L_{\rm X}/L_{\rm IR})$ of our sample.
(b) Relation between IR (8--1000 $\mu$m)
and AGN bolometric luminosities of our Hy/U/LIRGs. The symbols are the same as (a). 
The grey dashed line corresponds to the $L_{\rm IR} = L_{\rm AGN}$ relation.}
\label{fig:e3}
\end{figure*}

\begin{figure}
\epsscale{0.494}
\plotone{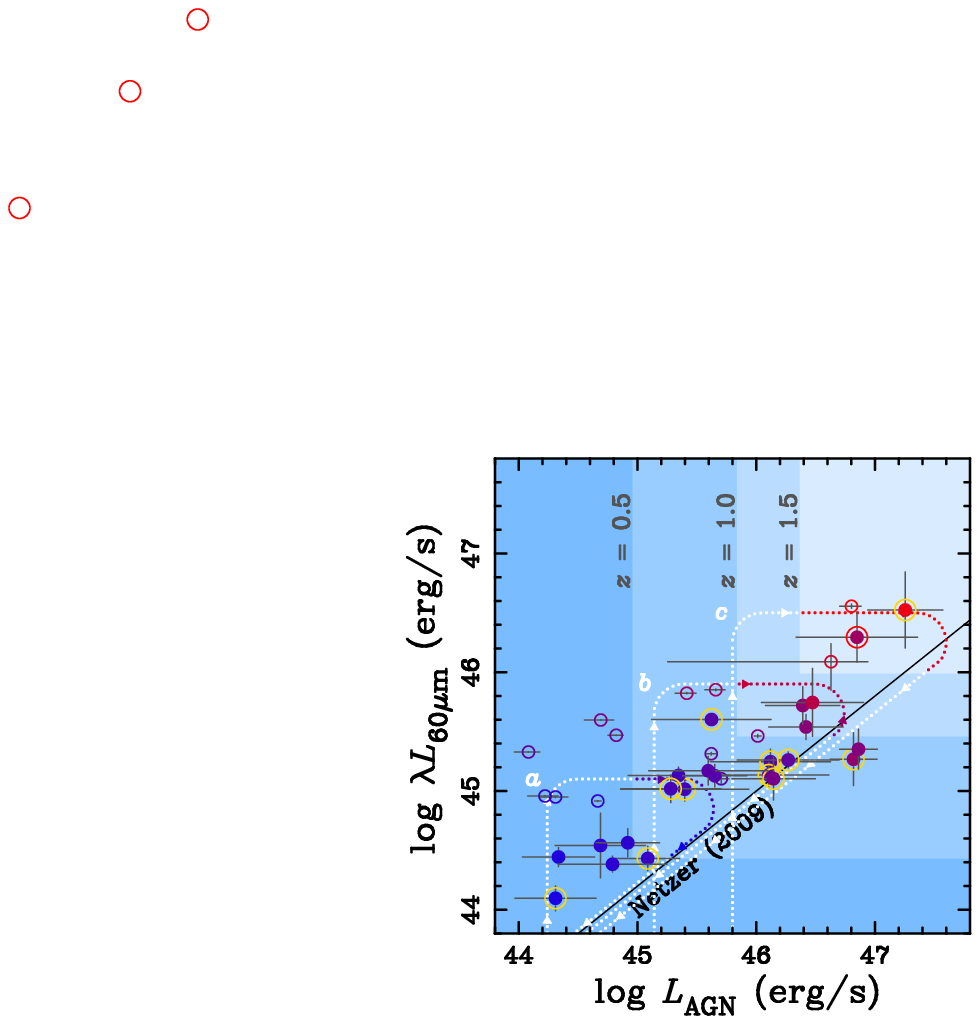}
\caption{Relation between 60 \micron\ and AGN bolometric luminosities of the 23 Hy/U/LIRGs, shown as filled circles color-coded according to the $\log (L_{\rm IR}/L_\odot)$ value.
Yellow-open circles indicate objects with $\log N_{\rm H} > 22$.
The black line denotes a relation for unobscured AGNs obtained by \citet{2009MNRAS.399.1907N}.
Three rectangle fields marked with $z = 0.5$, 1.0, and 1.5 denote the detectable area for objects with $N_{\rm H} = 10^{23}$ cm$^{-2}$ at each redshift above the \nustar\ and \spitzer\ detection limits.
Three dotted curves denote SF-AGN evolution sequences (see text).
Local U/LIRGs \citep{2014ApJ...794..139I} are shown with open circles color-coded according to IR luminosity.}
\label{fig:e4}
\end{figure}

Due to an error in converting \nustar\ count rate to flux, all intrinsic 2--10 keV luminosities ($L_{\rm X}$) are overestimated exactly by a factor of 2, and accordingly the bolometric AGN luminosities ($L_{\rm AGN}$) are also overestimated by similar factors.
Absorption column densities ($N_{\rm H}$) are correct.
Revised Table~\ref{tab:t} and Figures~\ref{fig:t} and \ref{fig:l} are presented below (i.e., Table~\ref{tab:e1} and Figures~\ref{fig:e3} and \ref{fig:e4}, respectively).

Our conclusions are little affected by this error.
In Figure~\ref{fig:t} (Sections~4.1 and 5), the X-ray to IR luminosity ratio of our sample is correctly $\log L_{\rm X}/L_{\rm IR} \sim -1.4$, which is larger than those of majority of local ULIRGs and the average value of PG QSOs.
In Figure~\ref{fig:l} (Sections~4.2 and 5), most of our objects are now located slightly above the \citet{2009MNRAS.399.1907N} line, but their $\lambda L_{60 \mu m}$ to $L_{\rm AGN}$ ratios are much smaller than those of SF-dominant IR-luminous galaxies reported by \citet{2014ApJ...794..139I}.
The implication that our X-ray detected Hy/U/LIRGs are likely in a transition phase between obscured and unobscured AGNs is unchanged.

\begin{deluxetable*}{cccccccc}
\tablecaption{Properties of 23 Hy/U/LIRGs \label{tab:e1}}
\tablewidth{0pt}
\tablecolumns{8}
\tablehead{
\colhead{ID$_{\it Sp}$\tablenotemark{$a$}} & \colhead{ID$_{\it Nu}$\tablenotemark{$b$}} &
\colhead{$z$\tablenotemark{$c$}} &
\colhead{$\log (L_{\rm IR}/L_\odot)$\tablenotemark{$d$}} &
\colhead{$\log \lambda L_{\rm 60{\mu}m}$\tablenotemark{$e$}} &
\colhead{$\log N_{\rm H}$\tablenotemark{$f$}} &
\colhead{$\log L_{\rm X}$\tablenotemark{$g$}} &
\colhead{$\log L_{\rm AGN}$\tablenotemark{$h$}}\\
\colhead{} & \colhead{} & \colhead{} & \colhead{} & \colhead{[erg s$^{-1}$]} &
\colhead{[cm$^{-2}$]} & \colhead{[erg s$^{-1}$]} & \colhead{[erg s$^{-1}$]}
}
\startdata
1190 & 103 & 0.520 & 11.83$\pm$0.04 & 45.01$\pm$0.10 & 23.4$_{-3.4}^{+0.5}$ & 43.91$\pm$0.38 & 45.40$\pm$0.54\\
1224 & 107 & 0.694 & 12.09$\pm$0.27 & 45.25$\pm$0.10 & 23.8$_{-1.0}^{+0.3}$ & 44.45$\pm$0.40 & 46.12$\pm$0.51\\
1222 & 111 & 0.500 & 11.65$\pm$0.22 & 45.13$\pm$0.07 & 20.0$_{-0.0}^{+0.0}$ & 43.87$\pm$0.11 & 45.35$\pm$0.42\\
1563 & 123 & 0.787 & 12.02$\pm$0.29 & 45.13$\pm$0.11 & 23.8$_{-1.4}^{+0.3}$ & 44.44$\pm$0.42 & 46.11$\pm$0.51\\
1928 & 145 & 0.445 & 11.29$\pm$0.26 & 44.43$\pm$0.10 & 22.2$_{-0.0}^{+0.1}$ & 43.67$\pm$0.07 & 45.09$\pm$0.27\\
2613 & 188 & 0.560 & 11.94$\pm$0.06 & 45.02$\pm$0.12 & 22.5$_{-0.0}^{+0.1}$ & 43.82$\pm$0.11 & 45.28$\pm$0.42\\
2659 & 192 & 0.360 & 11.61$\pm$0.26 & 44.54$\pm$0.27 & 20.9$_{-0.9}^{+0.3}$ & 43.36$\pm$0.10 & 44.69$\pm$0.38\\
2872 & 194 & 1.156 & 12.53$\pm$0.31 & 45.27$\pm$0.22 & 22.1$_{-0.1}^{+0.0}$ & 44.96$\pm$0.05 & 46.82$\pm$0.20\\
3064 & 206 & 1.024 & 12.36$\pm$0.27 & 45.35$\pm$0.17 & 21.9$_{-0.1}^{+0.0}$ & 44.99$\pm$0.04 & 46.86$\pm$0.16\\
3180 & 216 & 0.760 & 12.14$\pm$0.28 & 45.26$\pm$0.07 & 23.8$_{-0.8}^{+0.2}$ & 44.56$\pm$0.32 & 46.27$\pm$0.43\\
3603 & 232 & 0.931 & 11.98$\pm$0.23 & 45.10$\pm$0.18 & 23.4$_{-0.1}^{+0.1}$ & 44.47$\pm$0.10 & 46.15$\pm$0.35\\
5191 & 245 & 1.250 & 12.07$\pm$0.31 & 46.30$\pm$0.21 & 24.0$_{-4.0}^{+0.5}$ & 44.98$\pm$0.51 & 46.85$\pm$0.51\\
5074 & 251 & 1.066 & 12.75$\pm$0.29 & 45.54$\pm$0.11 & 20.0$_{-0.0}^{+0.0}$ & 44.67$\pm$0.08 & 46.42$\pm$0.31\\
5041 & 253 & 0.212 & 11.07$\pm$0.10 & 44.10$\pm$0.10 & 22.7$_{-0.0}^{+0.1}$ & 43.06$\pm$0.09 & 44.31$\pm$0.34\\
4644 & 256 & 0.260 & 11.38$\pm$0.03 & 44.44$\pm$0.08 & 20.6$_{-0.6}^{+0.6}$ & 43.08$\pm$0.08 & 44.33$\pm$0.30\\
4383 & 287 & 0.658 & 12.05$\pm$0.25 & 45.13$\pm$0.09 & 20.0$_{-0.0}^{+0.0}$ & 44.10$\pm$0.07 & 45.65$\pm$0.27\\
3756 & 296 & 1.108 & 12.36$\pm$0.23 & 45.72$\pm$0.16 & 21.5$_{-0.5}^{+0.3}$ & 44.65$\pm$0.08 & 46.39$\pm$0.31\\
4017 & 307 & 0.345 & 11.29$\pm$0.26 & 44.38$\pm$0.07 & 20.9$_{-0.3}^{+0.1}$ & 43.44$\pm$0.12 & 44.79$\pm$0.46\\
3967 & 311 & 0.688 & 12.37$\pm$0.03 & 45.60$\pm$0.04 & 22.9$_{-0.0}^{+0.1}$ & 44.08$\pm$0.13 & 45.62$\pm$0.50\\
3827 & 320 & 0.690 & 12.07$\pm$0.26 & 45.17$\pm$0.12 & 20.0$_{-0.0}^{+0.0}$ & 44.06$\pm$0.13 & 45.60$\pm$0.50\\
3654 & 322 & 0.356 & 11.45$\pm$0.04 & 44.56$\pm$0.12 & 21.9$_{-0.1}^{+0.0}$ & 43.54$\pm$0.07 & 44.92$\pm$0.27\\
3424 & 337 & 1.454 & 12.87$\pm$0.33 & 45.75$\pm$0.29 & 21.4$_{-1.4}^{+0.3}$ & 44.71$\pm$0.11 & 46.47$\pm$0.43\\
3638 & 339 & 1.845 & 13.03$\pm$0.09 & 46.52$\pm$0.32 & 23.3$_{-0.1}^{+0.0}$ & 45.27$\pm$0.08 & 47.26$\pm$0.32
\enddata
\tablecomments{
All the errors are 1$\sigma$.
}
\tablenotetext{$a$}{Identification number in the \spitzer-COSMOS catalog \citep{2010ApJ...709..572K}.}
\tablenotetext{$b$}{\nustar\ source number from \citet{2015ApJ...808..185C}.}
\tablenotetext{$c$}{Redshift provided by \citet{2010ApJ...709..572K}.}
\tablenotetext{$d$}{IR luminosity integrated from 8 \micron\ to 1000 \micron\ in \citet{2010ApJ...709..572K}.}
\tablenotetext{$e$}{FIR luminosity at 60 \micron\ (rest-frame) estimated by using \spitzer/MIPS data at 24, 70, and 160 \micron\ (observed frame) bands (see Section~4.2).}
\tablenotetext{$f$}{Absorption hydrogen column density estimated by using the HR value (see Section~3 for more details). We adopt $\log N_{\rm H} = 20.0$ as a dummy lower limit.}
\tablenotetext{$g$}{Intrinsic (de-absorbed) luminosity in the rest-frame 2--10 keV band.}
\tablenotetext{$h$}{Bolometric AGN luminosity converted from the 2--10 keV luminosity by adopting a conversion factors given in \citet{2009ApJ...700.1878R}.}
\end{deluxetable*}


\begin{thebibliography}{}
\bibitem[Aird et al.(2015)]{2015ApJ...815...66A} Aird, J., Alexander, D.~M., Ballantyne, D.~R., et al.\ 2015, \apj, 815, 66
\bibitem[Alexander \& Hickox(2012)]{2012NewAR..56...93A} Alexander, D.~M., \& Hickox, R.~C.\ 2012, \nar, 56, 93
\bibitem[Berta et al.(2003)]{2003A&A...403..119B} Berta, S., Fritz, J., Franceschini, A., Bressan, A., \& Pernechele, C.\ 2003, \aap, 403, 119
\bibitem[Blecha et al.(2011)]{2011MNRAS.412.2154B} Blecha, L., Cox, T.~J., Loeb, A., \& Hernquist, L.\ 2011, \mnras, 412, 2154
\bibitem[Brusa et al.(2007)]{2007ApJS..172..353B} Brusa, M., Zamorani, G., Comastri, A., et al.\ 2007, \apjs, 172, 353
\bibitem[Brusa et al.(2010)]{2010ApJ...716..348B} Brusa, M., Civano, F., Comastri, A., et al.\ 2010, \apj, 716, 348
\bibitem[Cappelluti et al.(2009)]{2009A&A...497..635C} Cappelluti, N., Brusa, M., Hasinger, G., et al.\ 2009, \aap, 497, 635
\bibitem[Chapman et al.(2000)]{2000MNRAS.319..318C} Chapman, S.~C., Scott, D., Steidel, C.~C., et al.\ 2000, \mnras, 319, 318
\bibitem[Civano et al.(2012)]{2012ApJS..201...30C} Civano, F., Elvis, M., Brusa, M., et al.\ 2012, \apjs, 201, 30
\bibitem[Civano et al.(2015)]{2015ApJ...808..185C} Civano, F., Hickox, R.~C., Puccetti, S., et al.\ 2015, \apj, 808, 185
\bibitem[Donley et al.(2012)]{2012ApJ...748..142D} Donley, J.~L., Koekemoer, A.~M., Brusa, M., et al.\ 2012, \apj, 748, 142
\bibitem[Elvis et al.(2009)]{2009ApJS..184..158E} Elvis, M., Civano, F., Vignali, C., et al.\ 2009, \apjs, 184, 158
\bibitem[Fiore et al.(2009)]{2009ApJ...693..447F} Fiore, F., Puccetti, S., Brusa, M., et al.\ 2009, \apj, 693, 447
\bibitem[Gandhi et al.(2009)]{2009A&A...502..457G} Gandhi, P., Horst, H., Smette, A., et al.\ 2009, \aap, 502, 457
\bibitem[G{\"u}ltekin et al.(2009)]{2009ApJ...698..198G} G{\"u}ltekin, K., Richstone, D.~O., Gebhardt, K., et al.\ 2009, \apj, 698, 198
\bibitem[Ho(2008)]{2008ARA&A..46..475H} Ho, L.~C.\ 2008, \araa, 46, 475
\bibitem[Hopkins \& Quataert(2010)]{2010MNRAS.407.1529H} Hopkins, P.~F., \& Quataert, E.\ 2010, \mnras, 407, 1529
\bibitem[Ichikawa et al.(2012)]{2012ApJ...754...45I} Ichikawa, K., Ueda, Y., Terashima, Y., et al.\ 2012, \apj, 754, 45
\bibitem[Ichikawa et al.(2014)]{2014ApJ...794..139I} Ichikawa, K., Imanishi, M., Ueda, Y., et al.\ 2014, \apj, 794, 139
\bibitem[Karouzos et al.(2014)]{2014ApJ...784..137K} Karouzos, M., Im, M., Trichas, M., et al.\ 2014, \apj, 784, 137
\bibitem[Kartaltepe et al.(2010)]{2010ApJ...709..572K} Kartaltepe, J.~S., Sanders, D.~B., Le Floc'h, E., et al.\ 2010, \apj, 709, 572
\bibitem[Kawamuro et al.(2016)]{2016ApJS..225...14K} Kawamuro, T., Ueda, Y., Tazaki, F., Ricci, C., \& Terashima, Y.\ 2016, \apjs, 225, 14
\bibitem[Lin et al.(2016)]{2016MNRAS.456.2735L} Lin, M.-Y., Hashimoto, Y., \& Foucaud, S.\ 2016, \mnras, 456, 2735
\bibitem[Magorrian et al.(1998)]{1998AJ....115.2285M} Magorrian, J., Tremaine, S., Richstone, D., et al.\ 1998, \aj, 115, 2285
\bibitem[Marconi \& Hunt(2003)]{2003ApJ...589L..21M} Marconi, A., \& Hunt, L.~K.\ 2003, \apjl, 589, L21
\bibitem[Matsuoka \& Woo(2015)]{2015ApJ...807...28M} Matsuoka, K., \& Woo, J.-H.\ 2015, \apj, 807, 28
\bibitem[Netzer et al.(2007)]{2007ApJ...666..806N} Netzer, H., Lutz, D., Schweitzer, M., et al.\ 2007, \apj, 666, 806
\bibitem[Netzer(2009)]{2009MNRAS.399.1907N} Netzer, H.\ 2009, \mnras, 399, 1907
\bibitem[Onaka et al.(2007)]{2007PASJ...59S.401O} Onaka, T., Matsuhara, H., Wada, T., et al.\ 2007, \pasj, 59, S401
\bibitem[Rieke et al.(2004)]{2004ApJS..154...25R} Rieke, G.~H., Young, E.~T., Engelbracht, C.~W., et al.\ 2004, \apjs, 154, 25
\bibitem[Rigby et al.(2009)]{2009ApJ...700.1878R} Rigby, J.~R., Diamond-Stanic, A.~M., \& Aniano, G.\ 2009, \apj, 700, 1878
\bibitem[Robertson et al.(2006)]{2006ApJ...641...90R} Robertson, B., Hernquist, L., Cox, T.~J., et al.\ 2006, \apj, 641, 90
\bibitem[Rosario et al.(2012)]{2012A&A...545A..45R} Rosario, D.~J., Santini, P., Lutz, D., et al.\ 2012, \aap, 545, A45
\bibitem[Rowan-Robinson et al.(1997)]{1997MNRAS.289..490R} Rowan-Robinson, M., Mann, R.~G., Oliver, S.~J., et al.\ 1997, \mnras, 289, 490
\bibitem[Sanders et al.(1988)]{1988ApJ...325...74S} Sanders, D.~B., Soifer, B.~T., Elias, J.~H., et al.\ 1988, \apj, 325, 74
\bibitem[Sanders et al.(2007)]{2007ApJS..172...86S} Sanders, D.~B., Salvato, M., Aussel, H., et al.\ 2007, \apjs, 172, 86
\bibitem[Santini et al.(2010)]{2010A&A...518L.154S} Santini, P., Maiolino, R., Magnelli, B., et al.\ 2010, \aap, 518, L154
\bibitem[Stalevski et al.(2016)]{2016MNRAS.458.2288S} Stalevski, M., Ricci, C., Ueda, Y., et al.\ 2016, \mnras, 458, 2288
\bibitem[Stanley et al.(2015)]{2015MNRAS.453..591S} Stanley, F., Harrison, C.~M., Alexander, D.~M., et al.\ 2015, \mnras, 453, 591
\bibitem[Teng \& Veilleux(2010)]{2010ApJ...725.1848T} Teng, S.~H., \& Veilleux, S.\ 2010, \apj, 725, 1848
\bibitem[Ueda et al.(2003)]{2003ApJ...598..886U} Ueda, Y., Akiyama, M., Ohta, K., \& Miyaji, T.\ 2003, \apj, 598, 886
\bibitem[Ueda et al.(2014)]{2014ApJ...786..104U} Ueda, Y., Akiyama, M., Hasinger, G., Miyaji, T., \& Watson, M.~G.\ 2014, \apj, 786, 104
\bibitem[Vega et al.(2008)]{2008A&A...484..631V} Vega, O., Clemens, M.~S., Bressan, A., et al.\ 2008, \aap, 484, 631
\bibitem[Veilleux et al.(1995)]{1995ApJS...98..171V} Veilleux, S., Kim, D.-C., Sanders, D.~B., Mazzarella, J.~M., \& Soifer, B.~T.\ 1995, \apjs, 98, 171
\bibitem[Veilleux et al.(1999)]{1999ApJ...522..113V} Veilleux, S., Kim, D.-C., \& Sanders, D.~B.\ 1999, \apj, 522, 113
\bibitem[Wilson et al.(2008)]{2008ApJS..178..189W} Wilson, C.~D., Petitpas, G.~R., Iono, D., et al.\ 2008, \apjs, 178, 189-224
\bibitem[Woo et al.(2013)]{2013ApJ...772...49W} Woo, J.-H., Schulze, A., Park, D., et al.\ 2013, \apj, 772, 49
\bibitem[Yuan et al.(2010)]{2010ApJ...709..884Y} Yuan, T.-T., Kewley, L.~J., \& Sanders, D.~B.\ 2010, \apj, 709, 884
\end{thebibliography}
\end{document}